\documentclass[aps,pre,twocolumn,superscriptaddress,showpacs]{revtex4-1}
\usepackage[dvips]{graphicx}
\usepackage{amsmath}
\begin{document}


\title{Noise-enhanced nonlinear response and the role of modular structure for signal detection in neuronal networks}



\author{M. A. Lopes}
\affiliation{Department of Physics $\&$ I3N, University of Aveiro, 3810-193 Aveiro, Portugal}
\author{K.-E. Lee}
\affiliation{Department of Physics $\&$ I3N, University of Aveiro, 3810-193 Aveiro, Portugal}
\author{A. V. Goltsev}
\affiliation{Department of Physics $\&$ I3N, University of Aveiro, 3810-193 Aveiro, Portugal}
\affiliation{A.F. Ioffe Physico-Technical Institue, 194021 St. Petersburg, Russia}
\author{J. F. F. Mendes}
\affiliation{Department of Physics $\&$ I3N, University of Aveiro, 3810-193 Aveiro, Portugal}





\begin{abstract}
We find that sensory noise delivered together with a weak periodic signal not only enhances nonlinear response of neuronal networks, but also improves the synchronization of the response to the signal. We reveal this phenomenon in neuronal networks that are in a dynamical state near a saddle-node bifurcation corresponding to appearance of sustained network oscillations.
In this state, even a weak periodic signal can evoke sharp nonlinear oscillations of neuronal activity. These sharp network oscillations have a deterministic form and amplitude determined by nonlinear dynamical equations. The signal-to-noise ratio reaches a maximum at an optimum level of sensory noise, manifesting stochastic resonance (SR) at the population level. We demonstrate SR by use of simulations and numerical integration of rate equations in a cortical model with stochastic neurons. Using this model, we mimic the experiments of Gluckman \emph{et al} [B. J. Gluckman et al, Phys. Rev. Lett. \textbf{77}, 4098 (1996)] that have given evidence of SR in mammalian brain.
We also study neuronal networks in which neurons are grouped in modules and every module works in the regime of SR. We find that even a few modules can strongly enhance the reliability of signal detection in comparison with the case when a modular organization is absent.
\end{abstract}


\maketitle

\section{Introduction \label{introduction}}

Noise is ubiquitous in sensory systems and strongly influences how they function \cite{Ermentrout_2008,Faisal_2008}.
Understanding how sensory systems compensate, counter or account for noise in order to detect and process sensory information remains elusive.
Stochastic resonance (SR) is recognized as a possible mechanism that allows sensory systems to use noise for its own benefit \cite{Ermentrout_2008,Faisal_2008,McDonnell_2011}.
SR is a phenomenon that describes an amplification
and an optimization of weak signals by noise. It was revealed in many physical systems \cite{Gammaitoni_1998}.
In the brain, SR was observed experimentally in sensory systems
\cite{Douglass_1993, Wiesenfeld_1995, Levin_1996,Russell_1999},
in central neurons such as hippocampal CA1 neurons in rat cortex \cite{Gluckman_1996,Stacey_2000,Stacey_2001}, in the human blood pressure
regulatory system \cite{Hidaka_2000}, and the human brain's visual
processing area \cite{Mori_2002}. SR is also considered as a mechanism mediating neuronal synchronization within and between functionally relevant brain areas \cite{Kitajo_2003,Ward_2009,Ward_2010}.

Most of the theoretical works on SR, including the seminal paper \cite{Benzi_1982},
and experimental realizations of SR refer to systems based on the
motion of a particle subjected to a weak periodic signal in a bistable (or multistable) potential \cite{Gammaitoni_1998}. In this case, the amplitude of the signal alone is insufficient to cause the particle to overcome the barrier between the wells. The addition of noise leads to a nonzero probability of transition from either well to the other which varies with the period of the signal. The transitions occur at random times but with some degree of correlation with the signal.
SR was also revealed in a class of dynamical systems based not on bistability but rather on
excitable dynamics near a saddle-node bifurcation \cite{Gang_1993,Wiesenfeld_1994}. Nonlinear dynamical systems belonging to this class consist only of a potential barrier (activation threshold), an applied (or intrinsic) weak periodic subthreshold signal, and noise.
A key ingredient of the system is the following property. If the system is kicked by a stimulus from its  `rest state' above an activation threshold, then it returns to the state deterministically, within a certain refractory time \cite{Gang_1993,Wiesenfeld_1994,Rappel_1994}.
In particular, nonlinear dynamical systems near a saddle-node bifurcation
can demonstrate this kind of excitability and, thus, they can be used as model systems for studying SR \cite{Gang_1993,Wiesenfeld_1994,Rappel_1994,Longtin_1997}. Note that this bifurcation is a mechanism of a dynamical transition into a state with low-frequency sustained oscillations. Based on these ideas, several single neuron models have been proposed to explain SR  observed in the brain \cite{Wiesenfeld_1994,Longtin_1997,Wiesenfeld_1995,Stacey_2000,Stacey_2001}. In these models, when noise is applied, the cell fires action potentials that are synchronized with subthreshold signals.
However, SR was observed not only at the single neuron level,
but also at the level of an entire sensory system, i.e., as a collective phenomenon.
Gluckman \emph{et al}  \cite{Gluckman_1996} revealed resonance in response of a neuronal network
from  mammalian brain on weak periodic electric stimuli at
a certain magnitude of the stochastic component of an electric field.
In the experiments,  in the presence of noise, weak periodic signals generated bursts of synchronous neuronal activity with some degree of correlation with the signal. This activity was not clearly seen at the single cell level.
Until now, no theoretical explanation of these experiments was proposed and understanding of SR at the population level remains elusive.
There are some studies of SR in arrays of neurons \cite{Collins_1995, McDonnell_2007} and summing
networks \cite{Chialvo_1997}. However, these approaches do not take into account interactions between neurons. SR has also been studied in small networks \cite{Perc_2007} in which neurons were modeled by a discrete map proposed by Rulkov \cite{Rulkov_2001}.
In addition, SR was found in a small group of interacting Hodgkin-Huxley neurons \cite{Gong_2005,Ozer_2009}.

\begin{figure}
\includegraphics[width=0.45\textwidth]{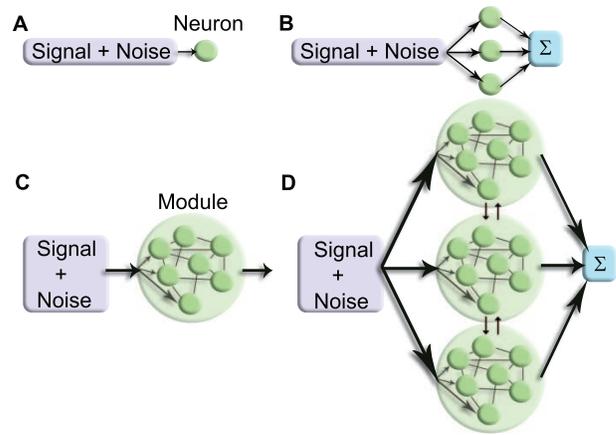}
\caption{(Color online) Different strategies of signal processing using
(A) a single neuron,
(B) an array of non-interacting single neurons,
(C) a random neuronal network, and
(D) a modular organization of a neuronal network.
\label{evolution-fig}}
\end{figure}

The experimental observations and theoretical investigations of SR both at the level of single neurons and at the level of neuronal populations revealed that, in the regime of SR, the response of nervous systems on weak signals still has a large stochastic component due to noise. It means that the detection of a signal is unreliable because a part of the input signal may be lost.
Therefore, the starting question remains: what does the system need in order to detect reliably
a sensory signal? At the present time it is well-recognized that network structure plays an important role in the function of nervous systems \cite{Bullmore_2009}. An important structural property is that,
in the brain, neurons of similar function are grouped together in columns (or modules). The columnar organization of the neocortex has been documented in studies of
sensory and motor areas in many species \cite{Mountcastle1955,HubelWiesel1962,Mountcastle2003}.
For example, rat somatosensory cortex has modular organization where neurons form columns and every column consists of 17000-19000 neurons (each rodent whisker has its own column) \cite{Meyer2010}. Even the sensory nervous system of the C. elegans, which is the smallest nervous system among animals, has modular organization \cite{Jarrell2012}.
At the present time, it is unclear what advantage, if any, is given by a modular organization.
The evolutionary origin of this structure was discussed by Kashtan and Alon \cite{Kashtan_2005}. They suggested that modularly varying goals leads to the spontaneous evolution of modular structure.
However, modular organization can have other advantages since it allows to perform parallel processing and subsequent summation and averaging of information
that are key principles used by sensory systems \cite{Faisal_2008}.
These ideas have been discussed in cognitive science within the connectionist model \cite{Feldman_1982}.
There are electrophysiological studies in monkeys which show that signals are averaged across neuronal modules and over time in the formation of a behavioral decision \cite{Gold_2000}.
If we assume that every module can work in the regime of SR, then this structure can easily overcome the limited reliability of SR in signal detection discussed above. Thus, nature may find in the modular organization  a way to introduce redundancy in order to increase reliability of signal detection.
One can speculate that this kind of modular organization may have appeared due to the process of evolution that resulted in the selection of such sensory systems that have higher reliability of signal detection. Sensory systems may have evolved from the processing of information by a single cell in Fig.~\ref{evolution-fig}(a) to arrays of cells in Fig.~\ref{evolution-fig}(b) and
from a neuronal network with a single module in Fig.~\ref{evolution-fig}(c)) to a modular organized neuronal network in Fig.~\ref{evolution-fig}(d).


In the present paper, at first we show that sensory noise delivered together with a weak periodic signal can not only enhance nonlinear response of neuronal networks, but also improves synchronization between the response and the signal.
We find this nonlinear phenomenon in neuronal networks that are in a dynamical state near a saddle-node bifurcation corresponding to appearance of sustained network oscillations.
This kind of excitable dynamics is similar to dynamics of the single neuron models discussed above in the context of SR \cite{Gang_1993,Wiesenfeld_1994,Rappel_1994,Longtin_1997}.
Using simulations of the cortical model \cite{Goltsev_2010,Lee_2014} with stochastic neurons and numerical integration of dynamical equations, we find that, when noise is applied, subthreshold sensory signals can generate activity of a large fraction of neurons. This activity has a form of sharp oscillations and is synchronized with some degree of correlation with the signal. The response
of neuronal activity on sensory signals and the signal-to-noise ratio (SNR)
reach a maximum at an optimum level of sensory noise. It manifests stochastic resonance at the population level. We mimic the experiments of Gluckman \emph{et al} \cite{Gluckman_1996} and find both qualitative and quantitative agreement with the data.
Second, we discuss the role of modular organization in the detection of weak signals. For this purpose, we study networks where neurons are grouped in modules and every module works in the regime of SR. We demonstrate that, in this case, the reliability of signal detection by the system is strongly enhanced in comparison with the case when a modular organization is absent.

\section{Cortical model \label{model}}

In the present section, we describe briefly properties of the cortical model with stochastic neurons, which we use to study SR in neuronal networks. The model was  introduced in \cite{Goltsev_2010} and generalized to the case of shot noise in  \cite{Lee_2014}. A similar model was proposed in \cite{Benayoun_2010,Wallace_2011}.

\subsection{Structure and rules of stochastic dynamics\label{structure}}
We consider neuronal networks composed of stochastic excitatory and inhibitory neurons.
The total number of neurons is $N$, the fraction of excitatory neurons is $g_e$, and the fraction of inhibitory neurons is $g_i=1-g_e$.
The neurons are connected by directed
edges (synapses) at random with the probability $c/N$ where
$c$ is the mean number of synaptic connections.
The network has a structure of the Erd\H{o}s-R\'{e}nyi
network with the Poisson degree distribution and small world
properties like the neuronal networks in the brain \cite{Sporns_2004}.
The neurons are bombarded by a flow of random delta-like spikes that represent spontaneous releases of neurotransmitters in synapses and random spikes arriving from other areas
of the brain. This noise has properties of shot noise. Neurons also receive spikes from active
presynaptic excitatory and inhibitory neurons.
Thus, the total input $I(t)$ at time $t$ to a neuron
is a sum of three contributions: (i) random spikes from shot
noise, (ii) spikes from excitatory neurons, and
(iii) spikes from inhibitory neurons.
The input $V_j$ to a neuron with index $j$, $j=1,2,\dots N$, is the integral of $I_{j}(t)$ over the time interval $[t-\tau,t]$,
\begin{equation}
V_j(t)= n J_{n} + kJ_e +lJ_i,
\label{input}
\end{equation}
where $n$, $k$, and $l$ are the numbers of spikes arriving during the time interval $[t-\tau,t]$ from shot noise, active presynaptic excitatory and inhibitory neurons, respectively.
$J_{n}$ is the amplitude of the shot noise spikes. $J_e$ and $J_i$ are the efficacies of synapses from excitatory and inhibitory neurons, respectively.

Dynamics of the cortical model with stochastic neurons is determined by the following rules. If during the integration time window $\tau$ the total input $V_j(t)$ to an inactive neuron becomes larger than a threshold value $\Omega$, then with the probability $\tau \mu_{a}$ the
neuron becomes active and fires a spike train with a constant frequency $\nu$
(the index $a=e$ if the neuron is excitatory and $a=i$ if it is inhibitory). If
the total input $V_j(t)$ of an active excitatory (inhibitory) neuron becomes smaller than $\Omega$, then the neuron stops to fire with the probability $\tau \mu_{a}$.
In this model, the rates $\mu_{e}$ and $\mu_{i}$ are the reciprocal
first-spike latencies  of excitatory and inhibitory neurons, respectively.
A stochastic behavior of neurons might be caused by an intrinsic noise within neurons \cite{Mainen_1995}, for example, by ion channel stochasticity \cite{Schneidman_1998}.

We introduce a parameter $\alpha$ that is the ratio of the first-spike latency of excitatory neurons to the first-spike latency of inhibitory neurons,
\begin{equation}
\alpha \equiv \mu_{i}/\mu_{e}.
\label{alpha}
\end{equation}
If $\alpha < 1$, then it means that excitatory neurons respond faster on stimuli than inhibitory neurons.

\subsection{Rate equations \label{rate equations}}
The fractions $\rho_e(t)$ and $\rho_i(t)$ of active excitatory and
inhibitory neurons, respectively, at time $t$
characterize neuronal activity.
They are determined by the following rate equations \cite{Goltsev_2010,Lee_2014}:
\begin{equation}
\frac{\dot{\rho_a}}{\mu_a}=-\rho_a+\Psi_a(\rho_e,\rho_i),
\label{rho-eq}
\end{equation}
where $a=e,i$, $\dot{\rho} \equiv d\rho /dt$. $\Psi_a(\rho_e,\rho_i)$ is the probability that, at time $t$, the total input to a randomly chosen excitatory ($a=e$) or inhibitory ($a=i$) neuron is at least the threshold $\Omega$.
The functions $\Psi_a(\rho_e,\rho_i)$ are determined by the network structure,
the distribution function of shot noise (we consider Gaussian
distribution for simplicity), and the frequency-current relationships
for single neurons (the step function in our model).
Note that the probability $\Psi_{a}(\rho_e,\rho_i)$ is the same for both excitatory and inhibitory neurons because,
in the network under consideration, excitatory and inhibitory neurons occupy topologically equivalent positions. Therefore, $\Psi_e(\rho_e,\rho_i)=\Psi_i(\rho_e,\rho_i)\equiv \Psi(\rho_e,\rho_i)$, where
\begin{eqnarray}
&\Psi(\rho_e,\rho_i)= \sum_{n,k,l\geq 0} \Theta(n J_{n}{+}J_ek{+}J_il{-}\Omega)G(n) \times \nonumber \\
& P_k(g_e\rho_e\tilde{c})P_l(g_i\rho_i\tilde{c}).
\label{psi-eq}
\end{eqnarray}
Here, $\tilde{c}=c \nu \tau$. $\Theta(x)$ is the Heaviside step function. $P_{k}(c)$ is the Poisson distribution function,
\begin{equation}
P_{k}(c)=c^{k}e^{-c}/k!
\end{equation}
$G(n)$ is the Gaussian distribution function,
\begin{equation}
G(n)=G_0 e^{-(n-\langle n \rangle)^2/2\sigma^2}.
\label{g-noise}
\end{equation}
$G(n)$ determines the probability that a neuron receives $n$ spikes from shot noise during the integration time $\tau$. $\langle n\rangle$ is the mean number of the spikes,
$\sigma$ is the variance,
and $G_0$ is the normalization constant, $\sum_{n=0}^{\infty}G(n)=1$. We use $\langle n\rangle$ as the control parameter characterizing the shot noise intensity.
Equations (\ref{rho-eq}) and (\ref{psi-eq}) are asymptotically exact in the thermodynamic limit, $N \rightarrow \infty$ \cite{Goltsev_2010,Lee_2014}.

In numerical simulations, we use the algorithm proposed in \cite{Lee_2014}.
We use the following model parameters: $N=10^4$,
$c=10^3$, $\Omega=30$, $\tau \nu = 1$, $\mu_{e} \tau =0.1$, $\alpha=0.7$, and $g_i=0.25$.
Throughout this paper we use $1/\mu_{e}\equiv 1$ as time unit and $J_e\equiv 1$ as input unit. Following \cite{Amit_1997}, we choose $J_{i}=-3J_{e}$. We also use $J_{n}=J_{e}$ and $\sigma^2 =10$ for the amplitude and the variance of shot noise. We use these parameters throughout the paper and consider only dependence of the network dynamics on the shot noise intensity $ \langle n \rangle$ and sensory signals.

\subsection{Phase diagram and sharp oscillations\label{phase diagram}}

\begin{figure}
\includegraphics[width=0.45\textwidth]{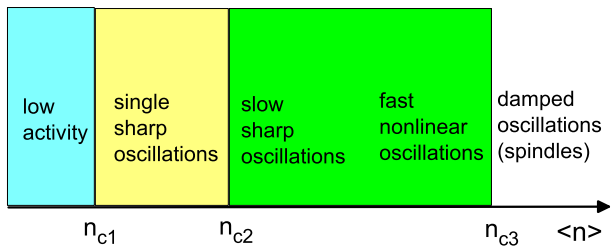}
\caption{(Color online) Phase diagram of the cortical model in dependence on shot noise intensity $\langle n \rangle$ in the case when excitatory neurons respond faster on input than inhibitory neurons ($\alpha=0.7$). With increasing $\langle n \rangle$, the neuronal network transits from region with low neuronal activity and weak fluctuations (the region $\langle n \rangle < n_{c1}$) into the region ($n_{c1} < \langle n \rangle < n_{c2}$) with single sharp oscillations (paroxysmal-like spikes of neuronal activity) that emerge irregularly from the low background activity.
Sustained network oscillations appear in the region  $n_{c2} < \langle n \rangle < n_{c3}$.
They have a large amplitude but a small frequency (slow sharp oscillations) above the critical point $n_{c2}$ of a saddle-node bifurcation in contrast to network oscillations near the Hopf bifurcation ($\langle n \rangle$ near $n_{c3}$) that have a small amplitude but a large frequency (fast oscillations). At $\langle n \rangle > n_{c3}$ the network oscillations are damped and fluctuations of neuronal activity appear in the form of spindle oscillations.
\label{phase_diagram-fig}}
\end{figure}

In this paper, we consider neuronal networks described by
the cortical model in the case when excitatory neurons respond faster on input than inhibitory neurons, i.e., $\alpha < 1$.
Figure \ref{phase_diagram-fig} displays the patterns of spontaneous neuronal activity in different regions of the shot noise intensity $ \langle n \rangle $ at $\alpha=0.7$ (adapted from \cite{Lee_2014}).
One can see that, if shot noise has a low intensity, $ \langle n \rangle < n_{c1}$, the neuronal network is in a state with a low neuronal activity and weak fluctuations around the state.
In the region $n_{c1} < \langle n \rangle < n_{c2}$,
the network still relaxes exponentially to the rest state if a perturbation
of neuronal activity (a simultaneous activation or inactivation of a certain number of neurons during a short time) is sufficiently weak.
However, if a perturbation is larger than an activation threshold, then strongly synchronized neuronal activity emerges in the form of single sharp oscillations with a large amplitude.
This activation threshold determines the number of excitatory neurons that must be activated simultaneously in order to generate a single sharp oscillation.
The duration of a single sharp oscillation is much larger than the period of spikes generated by single neurons.
At $\langle n \rangle > n_{c2}$, large-amplitude slow sustained network oscillations appear. Neuronal activity at $\langle n \rangle > n_{c2}$ is shown in Fig. \ref{phase_diagram-fig} and is explained in the caption of the figure.
Note that $\langle n \rangle = n_{c2}$ is the critical point of a second-order phase transition caused by a saddle-node bifurcation. The critical point $\langle n \rangle = n_{c1}$ is the end point of the region $n_{c1} < \langle n \rangle < n_{c2}$ in which the rate equations have three fixed points.

The single sharp oscillations (paroxysmal-like spikes) are collective events having interesting properties \cite{Lee_2014}. First, single sharp oscillations are deterministic and strongly nonlinear events that are formed by synchronous activity of almost $90 \%$ of neurons.
A single sharp oscillation is described by a trajectory (heteroclinic orbit) that goes around an unstable point in the $(\rho_e,\rho_i)-$phase plane.
Second, the activation threshold of a sharp oscillation depends on model parameters. It tends to zero when the network approaches the bifurcation critical point $n_{c2}$.
For example, in a network of $10^4$ neurons (7500 excitatory and 2500 inhibitory neurons), at $ \langle n \rangle= 16$, the simultaneous activation of 75 excitatory neurons chosen at random among 7500 excitatory neurons (i.e., about $1 \%$ of excitatory neurons), while the other neurons are inactive at that moment, generates a single sharp oscillation as a result of the synchronized activity of about 9000 neurons  \cite{Lee_2014}.
This phenomenon opens the possibility to observe SR in neuronal networks. When noise is applied, the neuronal network fires sharp oscillations with some degree of correlation
with subthreshold sensory signals.
This mechanism of SR is similar to the mechanism discussed within single neuron models \cite{Wiesenfeld_1994,Longtin_1997,Wiesenfeld_1995,Stacey_2000,Stacey_2001}. The important difference is that, in our model, SR is a collective phenomenon.
When $\langle n \rangle$
is close to $n_{c2}$, fluctuations of neuronal activity caused, for example, by finite-size effects can also generate the sharp oscillations and mask the useful response of the network on weak sensory signals. This effect restricts the region where SR may be observed in neuronal networks.

\section{Stochastic resonance in the cortical model \label{sr}}

In this section, we demonstrate SR in neuronal networks described by the cortical model.
In particular, we show that the model allows us to explain SR observed in mammalian brain
\cite{Gluckman_1996}. In the experiments \cite{Gluckman_1996}, hippocampal slices of rat's brain were stimulated by a time-varying electric field. The field had two components, a stochastic one representing
noise and the other representing a signal. As the magnitude of the stochastic component was increased, a resonance was observed in the response of the neuronal network to the weak periodic signal.

In order to explain these experiments, let us study the response of a neuronal network described by the cortical model on a weak periodic stimuli in the case when shot noise intensity is in the range $n_{c1} < \langle n \rangle < n_{c2}$ (see Fig. \ref{phase_diagram-fig}). We use  both numerical integration of rate equations (\ref{rho-eq}) and simulations.

In our numerical calculations and simulations, we assume that the first-spike latencies $1/\mu_e$ and $1/\mu_i$ of excitatory and inhibitory neurons equal to $20$ ms and $28.6$ ms, respectively. 
This choice of the first-spike latencies agrees with experimental data according to which the first spike latency is ranged from 25 to 49 ms for CA3 hippocampal pyramidal (excitatory) neurons \cite{fmi04} and from 20 to 128 ms for inhibitory cerebellar stellate cells \cite{mfmt05}.
For the chosen parameters, the cyclic frequency of sustained network oscillations in the cortical model is about $5.2$ Hz (when the noise intensity is $\langle n \rangle =25$ above  $n_{c2}\approx 18.8$). The frequency lies in the range of frequencies $4-12$ Hz of theta waves observed in the brain \cite{Bragin_1995}. The shape of our sustained network oscillations resembles theta waves measured by EEG in the hippocampus of rats (see Fig. 4 in \cite{Bragin_1995}). The burst frequency observed by Gluckman \emph{et al.} was also within the range $4-12$ Hz. The observed bursts (synchronous population events) typically lasted for $10-30$ ms. In our model, the single sharp spikes last for about $74$ ms at $\alpha=0.7$.

\subsection{Numerical integration}
\label{numerical}
\begin{figure}
\includegraphics[width=0.45\textwidth]{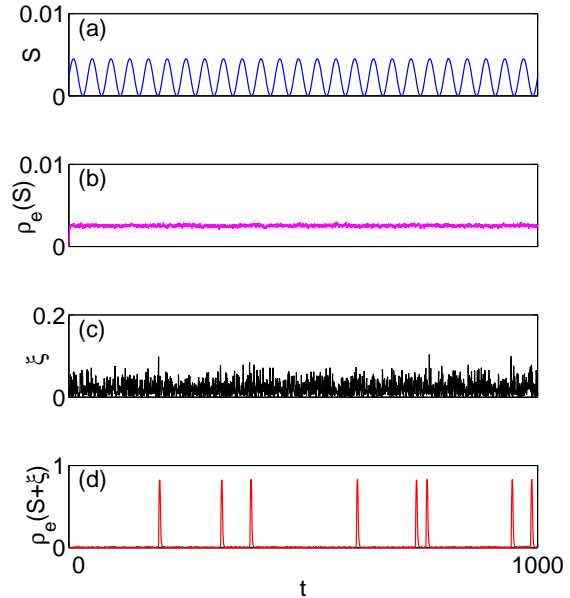}
\caption{(Color online)
In the absence of sensory noise, a periodic sensory signal ($S$) with the amplitude $A_s=4.5\times 10^{-3}$ (see the panel (a)) generates a weak perturbation of the excitatory population activity $\rho_e$ that can hardly be seen in the panel (b).
However, the addition of
sensory noise $\xi$ with the mean amplitude $A_{\xi}=2.15 \times 10^{-2}$ (see panel (c)), which is five times larger than the signal amplitude $A_s$, results in neuronal activity with sharp single oscillations shown in the panel (d).
The sharp single oscillations appear preferentially near the peaks of the sensory signal.
Network parameters: $c=1000$, $\Omega=30$, $g_i=0.25$, $J_{i}=-3J_{e}$, $\sigma^2=10$,
$\langle n\rangle=10$, $\alpha=0.7$, $g_s=0.1$, and $f_s=1.25$ Hz. Time $t$ is in units $1/\mu_e$.\label{sr_signal-fig}}
\end{figure}

First, we discuss results of the numerical integration of Eq.~(\ref{rho-eq}). We stimulate the neuronal network with a sensory stimulus $x(t)$ that contains both noise
$\xi(t)$ and a periodic signal $S(t)$,
\begin{equation}
x(t)=\xi(t)+S(t).
\label{s-input}
\end{equation}
We assume that the sensory stimulus is delivered
by $g_sN$ sensory neurons that we introduce in the cortical model in the following way. We connect these additional sensory neurons at random with the probability $c/N$ only to excitatory neurons. 
Therefore, each excitatory neuron receives input from, in average, $g_s c$ sensory neurons.
One can show that introduction of the sensory neurons leads to a simple modification of Eq.~(\ref{rho-eq}). Namely, in Eq.~(\ref{rho-eq}), we must substitute the function $\Psi(\rho_e,\rho_i)$ by $\Psi(\rho_e+A_{e}(t),\rho_i)$
where $A_{e}(t)=x(t)g_s/(g_e \nu \tau)$. We also introduce
an additional stochastic force $F(t)$ acting on neurons and
representing other sources of noise different from shot and sensory noise (for example, the force can represent irregular fluctuations caused by finite-size effects). Equation (\ref{rho-eq}) takes a form,
\begin{equation}
\frac{\dot{\rho_a}}{\mu_a}=(1-\rho_a)F(t)-\rho_a+\Psi(\rho_e+A_e(t),\rho_i).
\label{rho2-eq}
\end{equation}
We consider the sensory noise $\xi(t)$ generated by the Gaussian process with by the mean number $\langle \xi(t) \rangle =7 \times 10^{-3}$
of random spikes per the integration time
$\tau$ and the variance $\sigma_{sn}^2=7.3\times10^{-4}$ (in our numerical calculations
we only use the positive part of this Gaussian process and the effective mean amplitude of noise, $A_{\xi}$,
is $2.15 \times10^{-2}$) (see Fig. \ref{sr_signal-fig}(c)).
Furthermore, the sensory signal is chosen to be sinusoidal,
\begin{equation}
S(t)=A_s[\sin(2\pi f_st)+1]/2,
\label{s-signal}
\end{equation}
with the amplitude $A_s=4.5\times 10^{-3}$
and the frequency $f_s=1.25$ Hz. The ratio of the signal amplitude to the mean level of sensory noise
is close to the one used in \cite{Gluckman_1996}.
The stochastic force $F(t)$ representing finite-size effects is considered to be uniformly distributed in the interval $[0,0.009]$.

Analyzing the dynamics of the cortical model, we find that,
in the absence of a periodic signal,
the sensory noise produces occasionally sharp oscillations.
Adding a sinusoidal subthreshold sensory signal, which alone can not generate network oscillations (see Fig.~\ref{sr_signal-fig}(b)), we find that
sharp spikes appear preferentially near the maximums of the sinusoidal subthreshold signal (see Fig.~\ref{sr_signal-fig}(d)).
The network responds to about $0.32$ of the input signal.
This signal recognition probability is close to the probability $1/3$ found in \cite{Gluckman_1996}.

\begin{figure}
\includegraphics[width=0.45\textwidth]{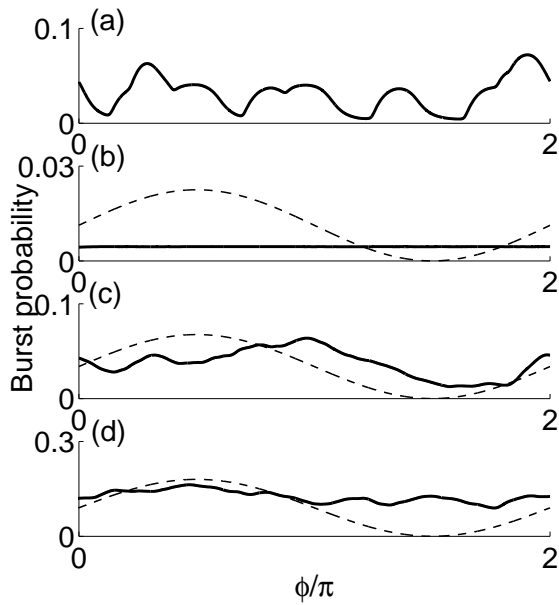}
\caption{Burst probability density (BPD) as a function of the phase $\phi$ of the sinusoidal signal.
(a) The BPD presents no modulation to noise alone as sensory input ($A_s=0$ and $A_{\xi}=0.0215$).
(b) An insufficient level of sensory noise does not generate a single burst ($A_s=4.5\times 10^{-3}$ and $A_{\xi}=6.1\times 10^{-3}$).
(c) An optimal level of sensory noise allows the neuronal network to recognize the weak input signal ($A_s=4.5\times 10^{-3}$ and
$A_{\xi}=0.0215$).
(d) Too strong sensory noise destroys the modulation ($A_s=4.5\times 10^{-3}$ and $A_{\xi}=0.0277$). The data are obtained by averaging over 250 periods of the signal.
The dashed lines represent the signal's phase.
Parameters in the numerical integration of the rate equations (\ref{rho2-eq}) are the same as in Fig. \ref{sr_burstprob-fig}.
\label{sr_burstprob-fig}}
\end{figure}

Following Gluckman \emph{et al.}, we also find the burst probability density (BPD) that is defined as the probability to observe a burst (a sharp spike in our case) of network activity  when the sinusoidal signal $S(t)$ has a phase $\phi$ (the signal maximums take place at $\phi=\pi (2n+1)/2$, where $n=0,1,\dots$ Figure~\ref{sr_burstprob-fig} displays  the BPD of the neuronal network
at different levels of sensory
noise. One can see that, the BPD correlates with the sensory signal (see Fig.~\ref{sr_burstprob-fig} (c)) around an optimal level of sensory noise, while no correlations were observed at  weaker or stronger levels of sensory noise (see Figs.~\ref{sr_burstprob-fig} (a) and  (d), respectively).
These results agree with the results of \cite{Gluckman_1996}.

\begin{figure}
\includegraphics[width=0.45\textwidth]{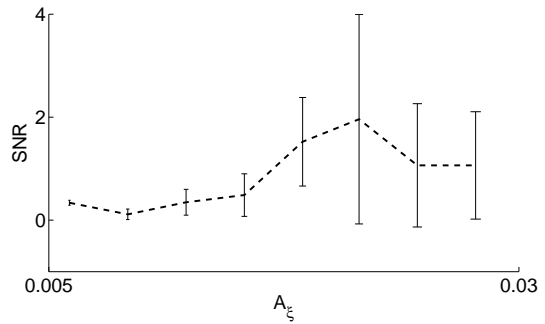}
\caption{SNR as function of the mean amplitude of sensory noise in the cortical model.
The inverted U-shape is characteristic of stochastic resonance.
Error bars were estimated from rms distribution of 10 measurements for
each level of noise.
Parameters in the numerical integration of the rate equations (\ref{rho2-eq}) are the same as in Fig. \ref{sr_burstprob-fig}.
\label{sr_snr-fig}}
\end{figure}

Finally, we find the signal-to-noise ratio (SNR) that we define as follows:
\begin{equation}
SNR=\frac{a}{b}
\label{snr-eq}
\end{equation}
where $a$ is the amplitude of the peak of the power spectral density (PSD) of neuronal activity at the signal's frequency $f_s$ and $b$ is the average value of the background PSD excluding the peak (Gluckman \emph{et al.} used a similar method).
We find the PSD by use of numerical integration of Eq.~(\ref{rho2-eq}) and numerical simulations of the cortical model. Within the methods, we applied the periodic sinusoidal signal in the presence of noise as discussed above and then analyzed the PSD of the neuronal activity.
Results of numerical integration of Eq.~(\ref{rho2-eq}) and estimation of the SNR for different levels of mean sensory noise are displayed in Fig.~\ref{sr_snr-fig}. The error bars represent the statistics: for each level of noise, we repeated $10$ times the measurements of the response of the neuronal network.
The maximum of the SNR that occurs at a nonzero level
of noise is a fingertip of stochastic resonance.

\subsection{Simulations}
\label{simulation}

\begin{figure}
\includegraphics[width=0.45\textwidth]{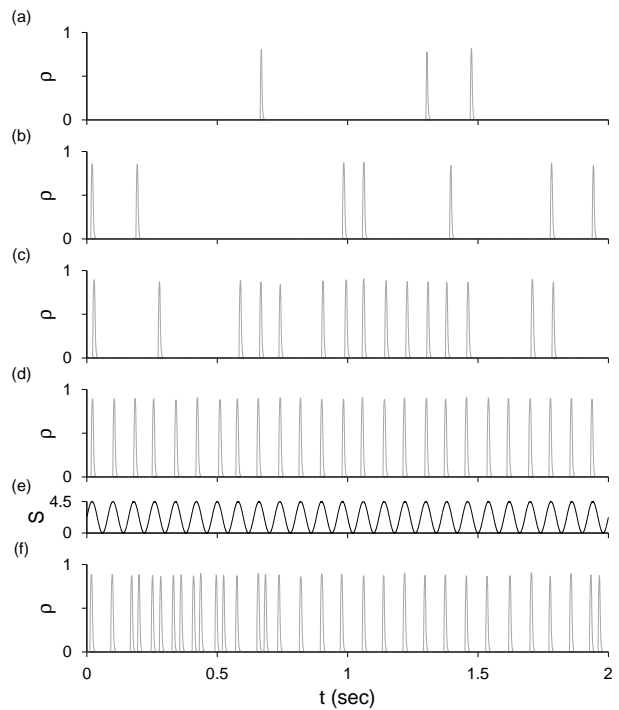}
\caption{Response of our neuronal network on sinusoidal sensory signal $S(t)$ at different levels  ($\langle \xi \rangle$) of sensory noise: (a) $\langle \xi \rangle =5.0$;
(b) $\langle \xi \rangle = 5.5$;
(c) $\langle \xi \rangle = 6.0$;
(d) $\langle \xi \rangle = 7.0$;
(e) sinusoidal signal $S(t)$;
(f) the response at $\langle \xi \rangle = 7.5$. Other parameters in simulations are the same as those in Fig. \ref{sr_burstprob-fig}.
\label{time_series}}
\end{figure}

In our simulations, in contrast to numerical calculations above, the sensory noise and sinusoidal signal (Eqs. (\ref{s-input}) and (\ref {s-signal})) were delivered directly to a fraction $g_s$ of excitatory neurons. Sensory noise $\xi$ was represented by random spikes with the mean number $\langle \xi \rangle$ of spikes per the integration time $\tau$ and the variance $\sigma_{sn}$.
The amplitude $A_s$ of the sinusoidal signal $S(t)$ was fixed. The level $\langle \xi \rangle$  of sensory noise was gradually increased. This enables us to study the impact of sensory noise on the response of the neuronal network.
Note that there is a simple relationship between $\langle \xi \rangle$ and the noise amplitude $A_{\xi}$ used in Sec. \ref{numerical}, $\langle \xi \rangle=c A_{\xi}$. In agreement with the numerical integration, we use the amplitude of the sinusoidal signal $A_s=4.5$. Other model parameters were the same as those in Sec. \ref{numerical}, except the parameters  $\langle \xi \rangle$ and the variance ($\sigma_{sn}^2=5$).

\begin{figure}
\includegraphics[width=0.3\textwidth]{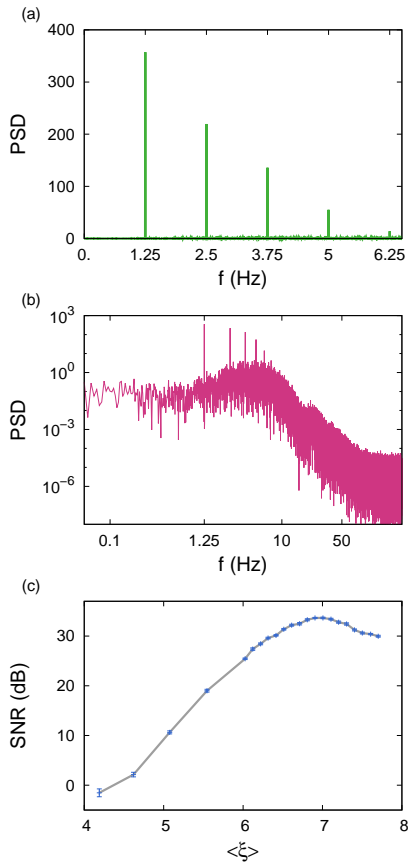}
\caption{Power spectral density (PSD) of the cortical model in which a small fraction ($g_s =0.1$) of excitatory neurons is stimulated by a sinusoidal signal (Eq. (\ref {s-signal})) with frequency $f_s=0.025$ (1.25 Hz) in the presence of sensory noise ($\langle \xi \rangle = 7.0$). (a) PSD vs. the frequency $f$, a linear scale; (b) PSD vs. $f$, a log-log scale. (d) Signal-to-noise (SNR) ratio vs. the mean amplitude of noise $\langle \xi \rangle$.
Parameters: $\alpha=0.7$, the intrinsic noise level $\langle n \rangle =10$, the amplitude of the sinusoidal signal $A_s=4.5$, the variance of sensory noise $\sigma_{sn}^2=5$, the integration time $\tau=0.1/\mu_e$. SNR is in decibel [$10 \log_{10} (SNR)$]. Other parameters in simulations are the same as in Fig. \ref{sr_burstprob-fig}.
\label{PSD_and_SNR}}
\end{figure}

At a small level $\langle \xi \rangle$ of sensory noise ($\langle \xi \rangle < 5$), response of the neuronal network on the sinusoidal sensory signal is weak since the probability of generation of sharp oscillations by the signal is small
(see Fig.~\ref{time_series}(a)). The sharp oscillations appear preferentially near the maximums of the signal $S(t)$. As the level $\langle \xi \rangle$ of sensory noise is increased, the response is enhanced and sharp oscillations are generated with a larger probability. Note also that the degree of correlation of the sharp oscillations with the sensory signal is also increased (compare Figs.~\ref{time_series}(b) and (c) with Fig.~\ref{time_series}(e)). At the optimum level of sensory noise ($\langle \xi \rangle \approx 7$), the network response (Fig.~\ref{time_series}(d)) is well synchronized with the sensory signal (Fig.~\ref{time_series}(e)). This noise-induced synchronization is amazing since only $10\%$ of excitatory neurons receive the signal+noise input despite the fact that the level of noise is larger than the signal amplitude. Finally, with increasing $\langle \xi \rangle$ above the optimum level, the synchronization become worse (see Fig.~\ref{time_series}(f)).

In order to characterize the network response, we also measured power spectral density of activity fluctuations and calculated the SNR from Eq.~(\ref{snr-eq}). Figures~\ref{PSD_and_SNR}(a) and (b) show the PSD of the neuronal activity displayed in Fig.~\ref{time_series}(d). One sees that the PSD has a strong peak at the frequency of the sinusoidal signal $S(t)$ (other peaks correspond to the respective harmonics). The value of this peak characterizes the network response. With increasing the level $\langle \xi \rangle$ of sensory noise, the peak becomes larger in comparison with the background amplitude of the PSD, and consequently the SNR increases (see Fig.~\ref{PSD_and_SNR}(c)). It means that the stronger the sensory noise, the large is the SNR and the better is the signal detection. Again, the inverted-U shape of the SNR is the hallmark of stochastic resonance.

Thus, numerical integration of Eq.~(\ref{rho2-eq}) and our simulations of the cortical model  show that sensory noise not only enhances nonlinear response of neuronal networks on a weak periodic signal, but also improves synchronization between the response and signal.
These results show that SR is an emergent property of neuronal networks that are in a dynamical state near a saddle-node bifurcation. The cortical model reproduces both qualitatively and quantitatively the experiments of  Gluckman \emph{et al.} \cite{Gluckman_1996}.



\section{Signal detection in a modular neuronal network}
\label{modular}

\begin{figure}
\includegraphics[width=0.45\textwidth]{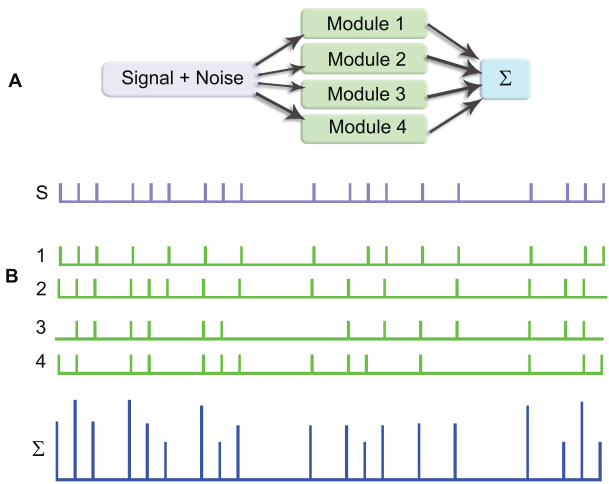}
\caption{(Color online) Detection of signals in networks with modular structure. (A) A signal together with
noise is sent to four modules 1, 2, 3 and 4. Responses of these modules are averaged in a
module denoted as $\Sigma$. (B) S represents the signal "ola" which is sent to these four modules.
This signal with noise generates random output signals 1, 2, 3 and 4 from the respective modules.
The signal $\Sigma$ represents the signal averaged over the output signals.
\label{message-fig}}
\end{figure}

The signal in Fig. \ref{sr_signal-fig}(a) carries no information.
Let us consider a case when a sensory signal
contains information. We choose the message "ola" ("hello" in Portuguese) expressed in Morse code as the digital code,
$1110111011100010111010100010111$. In order to represent this
message as a sensory signal, we consider rectangular pulses
separated by a time interval equal to 235 ms (the period of the sustained network
oscillations of 5.2 Hz).
The duration of these
pulses was chosen about $30$ ms that is about 8 times smaller than the period of network  oscillations.
The number of these pulses equals to the number of bits in our message. Finally, we
remove pulses corresponding to zeros. As a result we obtain a sensory signal
representing our message "ola" (see Fig. \ref{message-fig}). Despite the pulse amplitude was chosen sufficiently small, every pulse can generate with a certain probability a single sharp oscillation in the neuronal network. Figure \ref{message-fig} shows that the response of the neuronal network to this message is stochastic even at the optimal level of sensory noise.
On one hand, the network misses some pulses and does not detect them. On the other hand, it may elicit "false" responses.
For given model parameters, sensory noise level, and signal amplitude, we measured the probability $p$ that a pulse in the signal is detected, i.e., the pulse generates a single sharp oscillation. For the parameters chosen in our model and the signal amplitude $A_s=0.0135$, numerical integration of Eqs. \ref{rho-eq} gives the probability $p \approx 5/7$. Alternatively, one says that
two pulses of seven may be missed or may be "false".

As we have discussed in the Introduction, recognition of the message may be remarkably improved, if the neuronal network has a modular organization.
In order to show this, we now consider a neuronal network composed of $n$ modules that act in the regime of SR. Sensory neurons provide divergent input to
neurons in these modules. The modules receive the same signal
but are affected by independent sources of noise.
Such divergence was observed, for example, in auditory inner hair cells
\cite{Glowatzki_2002}. Then, responses of the modules to the sensory signal are combined and we obtain an averaged response shown in Fig.~\ref{message-fig}.
For every pulse in the message, the probability that at least one of the modules detects this pulse is
\begin{equation}
\Pi(n)=1-(1-p)^n.
\label{pr-n1}
\end{equation}
One can see that this probability increases with increasing $n$ as $\Pi(n) \approx  np$ at $p\ll 1$. In turn, the probability of an error, $1-\Pi(n)$, decreases exponentially with increasing $n$ as $1-\Pi(n)=\exp[-n |\ln(1-p)|]$. If we want to detect every pulse of the message
with probability of $99\%$, then the necessary number $n$ of
modules can be found from the condition $\Pi(n)=0.99$ (see,
for example, Ref. \cite{Carlson_2001}). For $p \approx 5/7$, this equation gives $n=4$. The response averaged over 4 neuronal modules on the message "ola" is shown in Fig. \ref{message-fig}. This result shows that modular structure with a few modules improves remarkably detection of weak signals.



Another advantage of the modular structure is that it increases the range of signal frequencies in which signal detection is reliable.
The frequency dependence of the probability $p$ follows from the fact that when the carrier frequency is about or larger than the reciprocal of the duration of a single sharp oscillation, then the processes of generation of single sharp oscillations by consecutive pulses strongly correlate with each other.
In our simulations of the cortical model, we measured the frequency dependence of the probability $p$ and found that $p$ decreases with increasing $f_s$. Decrease of $p$ below a certain value determines a range of frequencies where the reliable signal detection is possible.

In general, $p$ decreases when decreasing the modular size $N/n$ or when increasing the carrier frequency $f_s$. From Eq.~(\ref{pr-n1}) one sees that this results in decrease of $\Pi(n)$,  if $p$ decreases faster than $1/n$ at large $n$. Since at small $n$, the probability $\Pi(n)$ increases with increasing $n$, we conclude that $\Pi(n)$ has a maximum at an optimum number of modules. Finding the optimal number of modules
is an optimization problem, which may be solved if it is known how the probability $p$ depends on modular size $N/n$, the carrier frequency $f_s$ of the signal, and other parameters.
This simple consideration shows that, in principle, if the natural selection is based on optimization of signal detection then it can lead to modular organization with an optimum number of modules.

\section{Conclusion \label{conclusion}}
In conclusion, in this paper, we studied the response of neuronal networks on a periodic signal in the presence of sensory noise
and intrinsic shot noise represented by a flow of random spikes bombarding neurons.
Our simulations of the cortical model with stochastic neurons and numerical integration of corresponding rate equations revealed that sensory noise can not only enhance nonlinear response of neuronal networks, but can also improve synchronization of the response to the signal.
We demonstrated this noise-enhanced response in the case of neuronal networks that are in a dynamical state near a saddle-node bifurcation corresponding to appearance of sustained network oscillations.
In this state, even a weak sensory input delivered to only about $10\%$ of neurons can evoke a sharp single nonlinear oscillation of neuronal activity synchronized with some degree of correlation with the signal.
This sharp nonlinear oscillations are nonlinear events that represent a strongly synchronized activity of a large fraction of neurons ($90\%$ of neurons in our model) and have a deterministic shape determined by nonlinear dynamical equations.
The signal-to-noise ratio (SNR) reaches a maximum at an optimum level of sensory noise. It manifests stochastic resonance (SR) at the population level.
The proposed mechanism of SR in neuronal networks is similar to the mechanism of SR discussed previously within single neuron models \cite{Wiesenfeld_1994,Longtin_1997,Wiesenfeld_1995,Stacey_2000,Stacey_2001}.
We suggest that this kind of network response represented by a strongly synchronized activity of a large fraction of neurons may also play an important role in various mechanisms of signal processing in the brain. The fact that the sharp oscillations have a determinist form
and can be evoked by a few neurons may be of crucial importance not only for signal detection, but also for information transmission and communication between different areas of the brain. This mechanism enables a small group of neurons to control a large neuronal network.

Based on the proposed approach, we mimicked the experiments of Gluckman \emph{et al} \cite{Gluckman_1996} who observed SR in hippocampal slices from mammalian brain.
Results of our numerical analysis are in both qualitative and
quantitative agreement with the experiments.
The results also support the suggestion given by Gluckman
\emph{et al} that SR may enhance effects of weak hippocampal theta or
more widespread gamma oscillations within the brain.
In our model, SR is a consequence of
a neuronal network
being in a dynamical state near a saddle-node bifurcation, which is responsible for the emergence of sustained network oscillations with frequency in the range of theta waves (or higher frequencies, depending on parameters).
In this paper, we have focused on the theta range in order to compare with Gluckman experiments. However, similar results can be obtained with frequencies in the gamma range that is related with the function of sensory systems.
We would like to note that the fluctuation and response phenomena occurring near this kind of  saddle-node bifurcation
are universal properties of this kind of bifurcation and do not qualitatively depend on an underlying model.

We suggest that SR at the single neuron and population levels
can coexist and cooperate in order to improve the performance of signal detection in sensory systems.
At first stage, a sensory signal can activate a group of neurons working in regime of SR as it was proposed in \cite{Wiesenfeld_1994,Longtin_1997,Wiesenfeld_1995,Stacey_2000,Stacey_2001}. Then, at the second stage, the activated neurons can stimulate a synchronized activity of a finite fraction of the neuronal network in the form of non-linear sharp oscillations which we studied in this paper.

Since SR can not provide reliable signal detection, we also studied the role of modular organization of neuronal networks in this process.
For this purpose, we considered networks in which neurons are grouped in modules that work in the regime of SR.  We demonstrated that
even a few modules can strongly enhance the reliability of signal detection in comparison with the case when a modular organization is absent.
We tested our suggestion by use of numerical integrations of the cortical model.
One also notes that the modular organization also results in an increase of the range of signal frequencies in which a reliable signal detection is possible. Further detailed analysis of experimental data is necessary to confirm the mechanisms of the signal detection proposed in this paper.




\section{Acknowledgements}
This work was partially supported by FET IP Project MULTIPLEX 317532,
the PTDC projects SAU-NEU/ 103904/2008,  FIS/ 108476 /2008, MAT/ 114515 /2009, the project PEst-C / CTM / LA0025 / 2011, and the project "New Strategies Applied to Neuropathological Disorders," cofunded by QREN and EU.
K.~E.~L. and M.~A.~L. were supported by the FCT Grants No. SFRH/ BPD/ 71883/2010 and No. SFRH/ BD/ 68743 /2010.
\bibliography{Lopes_bib}
\end{document}